\documentclass[aps,prl,twocolumn,preprintnumbers,amsmath,amssymb,showpacs]{revtex4-1}
\usepackage{amsfonts}
\usepackage{bbm}
\usepackage{mathrsfs}
\usepackage{amsmath}
\usepackage{color}
\usepackage{graphicx}
\usepackage{epstopdf}
\usepackage{dcolumn}
\usepackage{bm}
\usepackage{longtable}
\usepackage{graphics}
\usepackage{amssymb}
\usepackage{xspace}
\usepackage{epsfig}
\usepackage{subfigure}
\usepackage{dcolumn}
\usepackage{multirow}
\makeatletter

\newcommand{\Rmnum}[1]{\expandafter\@slowromancap\romannumeral #1@}
\makeatother

\begin{document}
\title{Surface State Magnetization and Chiral Edge States on Topological Insulators}
\author{Fan Zhang}\email{zhf@sas.upenn.edu}
\author{C. L. Kane}
\author{E. J. Mele}
\affiliation{Department of Physics and Astronomy, University of Pennsylvania, Philadelphia, PA 19104, USA}
\begin{abstract}
We study the interaction between a ferromagnetically ordered medium and the surface states of a topological insulator with a general surface termination.
This interaction is strongly crystal face dependent and can generate chiral states along edges between crystal facets even for a uniform magnetization.
While magnetization parallel to quintuple layers shifts the momentum of Dirac point,
perpendicular magnetization lifts the Kramers degeneracy at any Dirac points {\em except} on the side face where the spectrum remains gapless and the Hall conductivity switches sign.
Chiral states can be found at any edge that {\em reverses} the projection of surface normal to the stacking direction of quintuple layers.
Magnetization also weakly hybridizes non cleavage surfaces.
\end{abstract}
\date{\today}
\maketitle

{\color{cyan}{\indent{\em Introduction.}}}---
Since the discovery of topological insulators (TI) \cite{Fu_Kane_Mele,Moore_Balents,Roy,Kane_RMP,Zhang_RMP} and the synthesis of real materials that realize their physics
\cite{BiSb,Zhang_DFT,BiSe,BiTe} there has been tremendous interest in their topologically protected surface states. Previous work has
focused mainly on the cleavage surfaces of ${\rm Bi_2Se_3}$, or similar TI's with $R \bar 3 m$ symmetry, that host spin-momentum locked helical metals with the conduction and valence bands
exhibiting opposite helicities. Interestingly, a quantum anomalous Hall (QAH) effect can be induced by exchange coupling the surface electrons to a magnetic insulator which lifts the Kramers degeneracy at the surface Dirac point (DP). When the magnetization is perpendicular to the quintuple layers, this introduces a mass term into the cleavage
surface state Hamiltonian, and if the Fermi energy is in this gap, there is a half integer quantized Hall conductivity $\sigma_{\rm H}=e^2/2h$ whose sign is determined by the direction of perpendicular magnetization. Theory predicts a 1D chiral edge state on the ${\rm Bi_2Se_3}$ cleavage surface along a domain wall where the perpendicular magnetization reverses direction \cite{Fu_Kane_Mele,Kane_RMP,TFT}. The fabrication of such an interface that displays the QAH effect poses a formidable experimental challenge.

In this work, we consider the effects of magnetic exchange coupling to topological surface states for a {\em general} crystal termination and discover new geometries that generically host 1D chiral edge channels. By breaking $\mathcal{T}$ symmetry the surface magnetization: (i) shifts the DP off $\mathcal{T}$ invariant momenta, (ii) couples non cleavage surfaces, and (iii) lifts the Kramers degeneracy at any DP {\em except} on the side face where the Hall conductivity switches sign. We find that all three effects are crystal face-dependent. Surprisingly, 1D gapless
chiral states can be induced at crystal edges {\em without} introducing a magnetic domain wall, accessing the
QAH effect in a geometry that should be readily accessible to experiment.
Interestingly, a recent experiment demonstrate bulk intergrowth of ${\rm Bi_2Se_3}$ and the room temperature ferromagnet ${\rm Fe_7Se_8}$ forms a ``stack of cards" structure \cite{Cava} that offers an opportunity for exploring the face-dependent interactions between TI surface states and ferromagnetic materials. Another advance in magnetically doped TI's realizes a $\sim 40$ meV gap \cite{dop2,Samarth} and a giant AH effect \cite{dop4} on the cleavage surface, providing a large out-of-plane Zeeman field to engineer the QAH effect in our new geometries.

\begin{figure}[t]
\centering{ \scalebox{0.4} {\includegraphics*{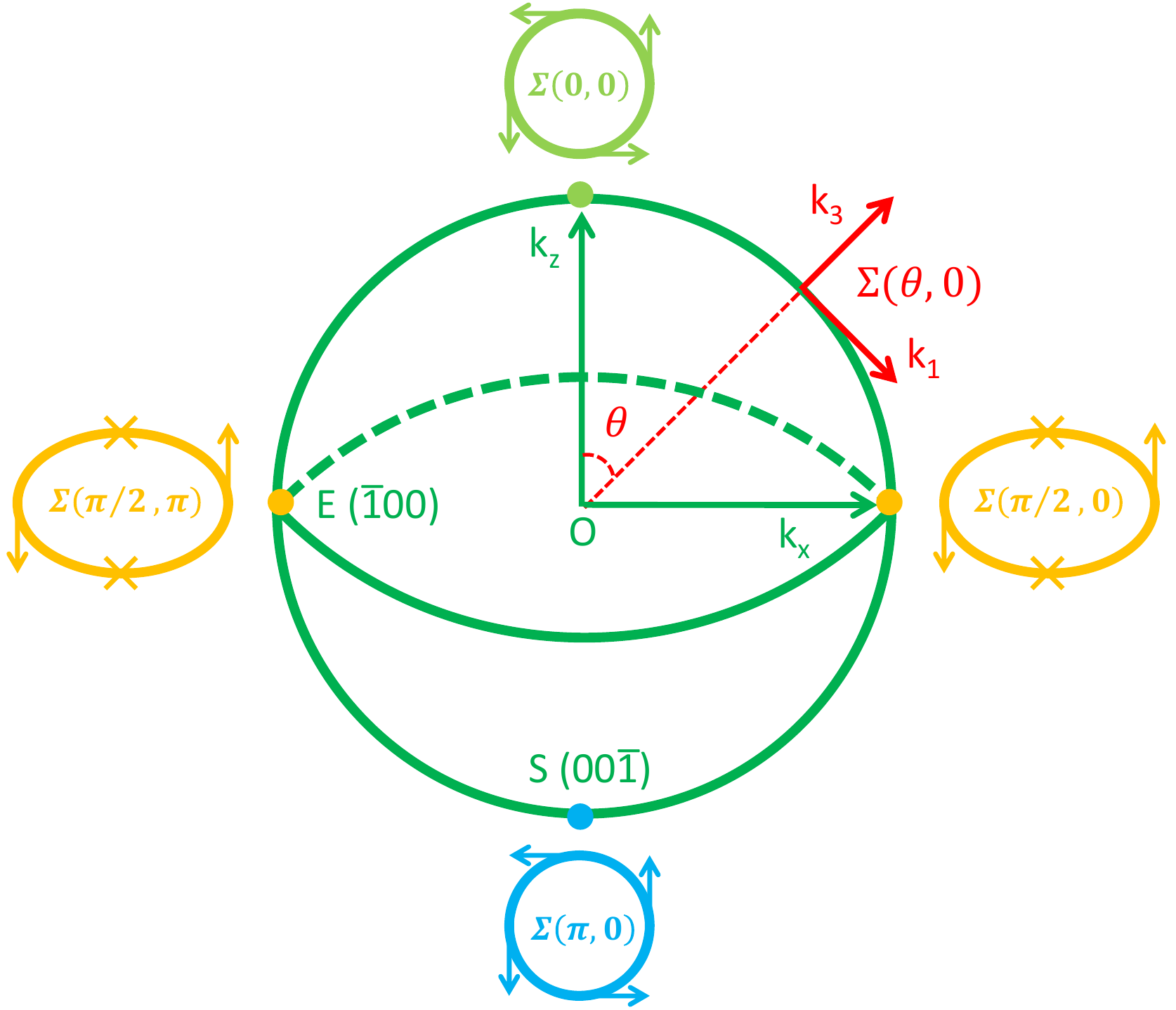}}}
\caption{\label{fig:fg1} {(color online). The definition of crystal and local frames and the sketch of the surface state conduction
band {\em spin} texture. $\hat{k}_{\rm z}$ is perpendicular to QL's. $\hat{k}_3$ is outward normal to the face $\Sigma(\theta,\phi)$ and $\hat{k}_1$ ($\hat{k}_2$) is in-plane
tangent to the longitude (latitude) circle with $\hat{k}_2=\hat{k}_3\times\hat{k}_1$.
The spin textures are shown in the local frames with $\hat{k}_1$ right and $\hat{k}_2$ up. The spin texture near DP is helical
on cleavage surfaces, filtered into $\pm \hat{k}_2$, maximized at $k_2=0$, and vanishing at $k_1=0$ on any side face, and tilted
out-of-plane otherwise.}}
\end{figure}

{\color{cyan}{\indent{\em Topological surface states.}}}---
We start from a description of the low energy minimal model of
${\rm Bi_2Se_3}$, followed by a derivation of the effective Hamiltonian of topological surface states near the DP of an arbitral face \cite{TISS}. These apply
generally to other TI's with $R\bar{3}m$ symmetry. Besides $\mathcal{T}$ and the parity
inversion ($\mathcal{P}$) symmetries, ${\rm Bi_2Se_3}$ crystal structure has threefold rotational ($\mathcal{C}_3$) symmetry
along $\hat{z}$ perpendicular to the quintuple layers (QL's), and twofold rotational ($\mathcal{C}_2$) symmetry along $\Gamma M$
direction. By convention we choose \cite{notation} the parity operator $\mathcal{P}=\tau_{\rm z}$ and the time reversal operator
$\mathcal{T}=iK\sigma_{\rm y}$ where $K$ is the complex conjugate operation. To linear order in $k$ the ${\bm k}\cdot{\bm p}$
bulk Hamiltonian that preserves the above four symmetries has a unique form
\begin{eqnarray}
\label{eqn:H}
\mathcal{H}_{\rm bulk}=-m\tau_{\rm z}+v_{\rm z}k_{\rm z}\tau_{\rm y}+v_{\shortparallel}(k_{\rm y}\sigma_{\rm x}-k_{\rm x}\sigma_{\rm y})\tau_{\rm x}\,,
\end{eqnarray}
where we assume $v_{\rm z}, v_{\shortparallel}>0$ and $\hbar=1$. By matching the eigensystems of TI with $m>0$ and
vacuum with $m\rightarrow -\infty$, one can demonstrate the existence of topological surface states \cite{TISS}. As shown in
Ref.\onlinecite{TISS}, the DP solution is determined by the operators ${\bm S_1}$ and is free under any
rotation of the operators ${\bm S_2}$. We first focus on the case with a crystal termination where the
azimuthal angle is fixed ($\phi=0$) and later generalize to situations where $\phi$ is allowed to vary between neighboring
crystal facets. For an arbitrary face $\Sigma(\theta)$ $(0\leq\theta\leq\pi)$ in Fig.\ref{fig:fg1}, ${\bm S_1}$ and ${\bm S_2}$
pseudospins read
\begin{eqnarray}
\label{eqn:structure}
{\bm S}_1&=\{\alpha\tau_{\rm x}+\beta\sigma_{\rm y}\tau_{\rm y},\alpha\tau_{\rm y}-\beta\sigma_{\rm y}\tau_{\rm x},\tau_{\rm z}\}\,,&\nonumber\\
{\bm S}_2&=\{\alpha\sigma_{\rm x}-\beta\sigma_{\rm z}\tau_{\rm z},\sigma_{\rm y},\alpha\sigma_{\rm z}+\beta\sigma_{\rm x}\tau_{\rm z}\}\,,&
\end{eqnarray}
where $v_3=\sqrt{(v_{\rm z}\cos\theta)^2+(v_{\shortparallel}\sin\theta)^2}$, $\alpha=v_{\rm z}\cos\theta/v_3$ and
$\beta=v_{\shortparallel}\sin\theta/v_3$. These pseudospins satisfy $[S_{\rm a}^{\rm i},S_{\rm b}^{\rm j}]=2i\delta_{\rm
ab}\epsilon^{\rm ijk}S_{\rm a}^{\rm k}$. We derive \cite{TISS} the topological surface state Hamiltonian for face
$\Sigma(\theta)$ to the linear order near DP,
\begin{eqnarray}
\label{eqn:Hs}
\mathcal{H}(\theta)=v_1 k_1\,S_2^{\rm y}-v_{\shortparallel}k_{\rm y}\,S_2^{\rm x}\,,
\end{eqnarray}
where $v_1=v_{\rm z}v_{\shortparallel}/v_3$. The surface band is the negative eigenstate of $S^{\rm x}_1$ and its chiral counterpart is localized on the opposite face. Thus the
surface state Hilbert space is reduced by half. The pseudospin (${\bm S}_2$) texture of surface states on a general face $\Sigma(\theta)$ is topologically equivalent to the helical metal found on the cleavage surface though its energy dispersion is anisotropic in momentum. However, the {\em spin} texture is different for each face \cite{TISS} and is determined by the bulk symmetries. Near the DP the spin texture of a constant energy contour is helical only on the top ($\theta=0$) and bottom ($\theta=\pi$) cleavage surfaces, it is collapsed to one dimension on any side face ($\theta=\pi/2$), and it is tilted out-of-plane otherwise. Interestingly on a side face with normal $\hat k_x$, the spin texture is filtered into $\pm \hat{k}_{\rm y}$ polarizations maximized at $k_{\rm y}=0$ and vanishing at $k_{\rm z}=0$. As a consequence, a Zeeman exchange field that couples to the physical spin $\sigma$ plays qualitatively different roles on different crystal faces.

Physically, a mass term $\Delta S_2^{\rm z}$ or $\Delta S_2^{\rm z}S_1^{\rm x}$ in Eq.(\ref{eqn:Hs}) is required to open an
energy gap at the surface DP. This amounts to introducing either one of the following external perturbations that break
$\mathcal{T}$ symmetry
\begin{eqnarray}
\label{eqn:Hgap}
\mathcal{H}^{\rm \Delta}_1=\Delta_1(\alpha\sigma_{\rm z}+\beta\sigma_{\rm x}\tau_{\rm z})\,,\quad\quad
\mathcal{H}^{\rm \Delta}_2=\Delta_2\sigma_{\rm z}\tau_{\rm x}\,.
\end{eqnarray}
The $\mathcal{H}^{\rm \Delta}_1$ terms depend on the surface orientation through the $\theta$ dependence of $\alpha$ and
$\beta$. On the cleavage surface ($\beta=0$) this perturbation is a Zeeman term that can be induced by an exchange field, while
on the side face ($\alpha=0$) it becomes $\sigma_{\rm x}\tau_{\rm z}$ which is negligibly small \cite{g-factor} since it
originates from the difference between electron spin ${\rm g}$-factors of ${\rm Bi}$ and ${\rm Se}$ \cite{notation}. In contrast,
$\mathcal{H}^{\rm \Delta}_2$ is independent of the crystal face angle but it requires a parity breaking interaction $\tau_{\rm
x}$ which seems to be infeasible.

\begin{figure}[b]
\centering{ \scalebox{0.48} {\includegraphics*{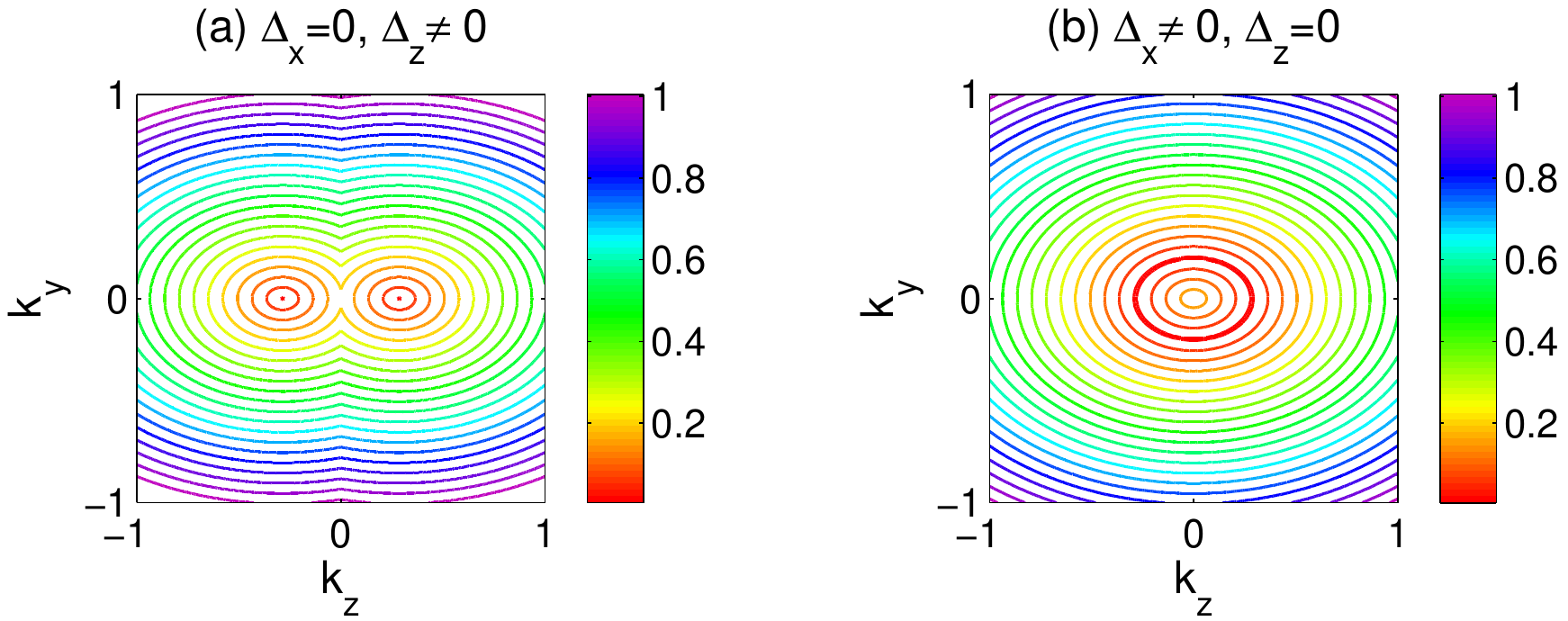}}}
\caption{\label{fig:fg2} {(color online). Constant energy contour plot of the lowest conduction band of the hybridized
surface states on two opposite side faces. (a) $\Delta_{\rm x}=0$, $\Delta_{\rm z}=0.2$; (b) $\Delta_{\rm x}=0.2$, $\Delta_{\rm z}=0$.
We use arbitrary units with  $v_{\shortparallel}=1$ and $v_{\rm z}=0.7$. There are two nodes in (a) but a zero-energy ellipse in (b).}}
\end{figure}

{\color{cyan}{\indent{\em Surface magnetization effects.}}}--- Now we consider a magnetic thin film
with uniform magnetization that provides an exchange coupling ${\bm \Delta}\cdot{\bm \sigma}$ to the spin of TI surface
states. Whether this breaking $\mathcal{T}$ symmetry coupling opens a gap at the DP on face $\Sigma(\theta)$ is
determined by whether it generates any of the perturbations listed in Eq.(\ref{eqn:Hgap}). Rewritten in the basis represented
by ${\bm S_1}$ and ${\bm S_2}$, the exchange coupling can be decomposed as follows:
\begin{eqnarray}
\label{eqn:Hx}
\Delta_{\rm x}\sigma_{\rm x}&=&\alpha\Delta_{\rm x} S_2^{\rm x}+\beta\Delta_{\rm x} S_2^{\rm z} S_1^{\rm z}\,,\\
\label{eqn:Hy}
\Delta_{\rm y}\sigma_{\rm y}&=&\Delta_{\rm y}S_2^{\rm y}\,,\\
\label{eqn:Hz}
\Delta_{\rm z}\sigma_{\rm z}&=&\alpha\Delta_{\rm z} S_2^{\rm z}-\beta\Delta_{\rm z} S_2^{\rm x} S_1^{\rm z}\,.
\end{eqnarray}
A topological surface state must be the negative eigenstate of $S^{\rm x}_1$ and its
positive counterpart is localized on the opposite face. Thus the two fields proportional to $S_1^{\rm z}$ in Eq.(\ref{eqn:Hx}) and Eq.(\ref{eqn:Hz}) couple the surface states on different
non cleavage faces ($\beta\neq 0$).
Although these two couplings play negligible roles as the
TI dimension becomes larger than the surface state penetration length they can be important for a sufficiently thin TI with two
parallel side faces. As illustrated in Fig.\ref{fig:fg2}, these two couplings hybridize the opposite surface states without opening any energy gap. The $S_2^{\rm x}
S_1^{\rm z}$ term breaks up the two DP's which repel each other in the $\hat{k}_1$ direction, leading to two zero
energy nodes at $v_{\rm z} k_{\rm z}=\pm\Delta_{\rm z}$ and $k_{\rm y}=0$ \cite{overlap}. $S_2^{\rm z} S_1^{\rm z}$ field splits the two DP's
in energy, resulting in a zero energy ellipse at $v_{\shortparallel}^2 k_{\rm y}^2+v_{\rm z}^2 k_{\rm z}^2=\Delta_{\rm
x}^2$ \cite{overlap}.

For magnetization parallel to QL's, the induced exchange terms $\Delta_{\rm x}S_2^{\rm x}$ and $\Delta_{\rm y}S_2^{\rm y}$ do not contain any mass terms listed in
Eq.(\ref{eqn:Hgap}), and thus not open gaps at any surface DP's. Instead, they break $\mathcal{T}$ symmetry by
shifting the DP from $\bar\Gamma$ to a non $\mathcal{T}$ invariant momentum
\begin{eqnarray}
\label{eqn:Hdp}
k_1=-\frac{\Delta_{\rm y}}{v_1}\,,\quad\quad
k_{\rm y}=\frac{\alpha\Delta_{\rm x}}{v_{\shortparallel}}\,.
\end{eqnarray}
Eq.(\ref{eqn:Hdp}) implies that magnetization only moves the DP on the side face in $\pm \hat{k}_{\rm z}$ direction. More
generally, the surface state velocities (or helicities) are opposite for opposite faces. Thus the same magnetization moves
their DP's in opposite directions in the crystal frame. Since a pair of opposite side faces are also connected by the
rotational symmetry along $\hat{k}_{\rm z}$, shifting the side face DP's is allowed along $\hat{k}_{\rm z}$ and forbidden along
$\hat{k}_2$ defined in Fig.\ref{fig:fg1}.

As shown in Eq.(\ref{eqn:Hz}), the magnetization perpendicular to QL's  introduces a field $\alpha\Delta_{\rm z} S_2^{\rm z}$ that
behaves as $\mathcal{H}^{\rm \Delta}_1$. This mass breaks $\mathcal{T}$ symmetry by lifting the Kramers degeneracy, leading
to a surface state gap $\sim 2\alpha\Delta_{\rm z}$. Importantly, on the closed surface of a compact TI this gap is
face-dependent: it is largest on the cleavage surface ($\theta = 0, \pi$) and it {\em vanishes} on the side face ($\theta=\pi/2$) where the mass switches sign.
\begin{figure}[t]
\centering{ \scalebox{0.36} {\includegraphics*{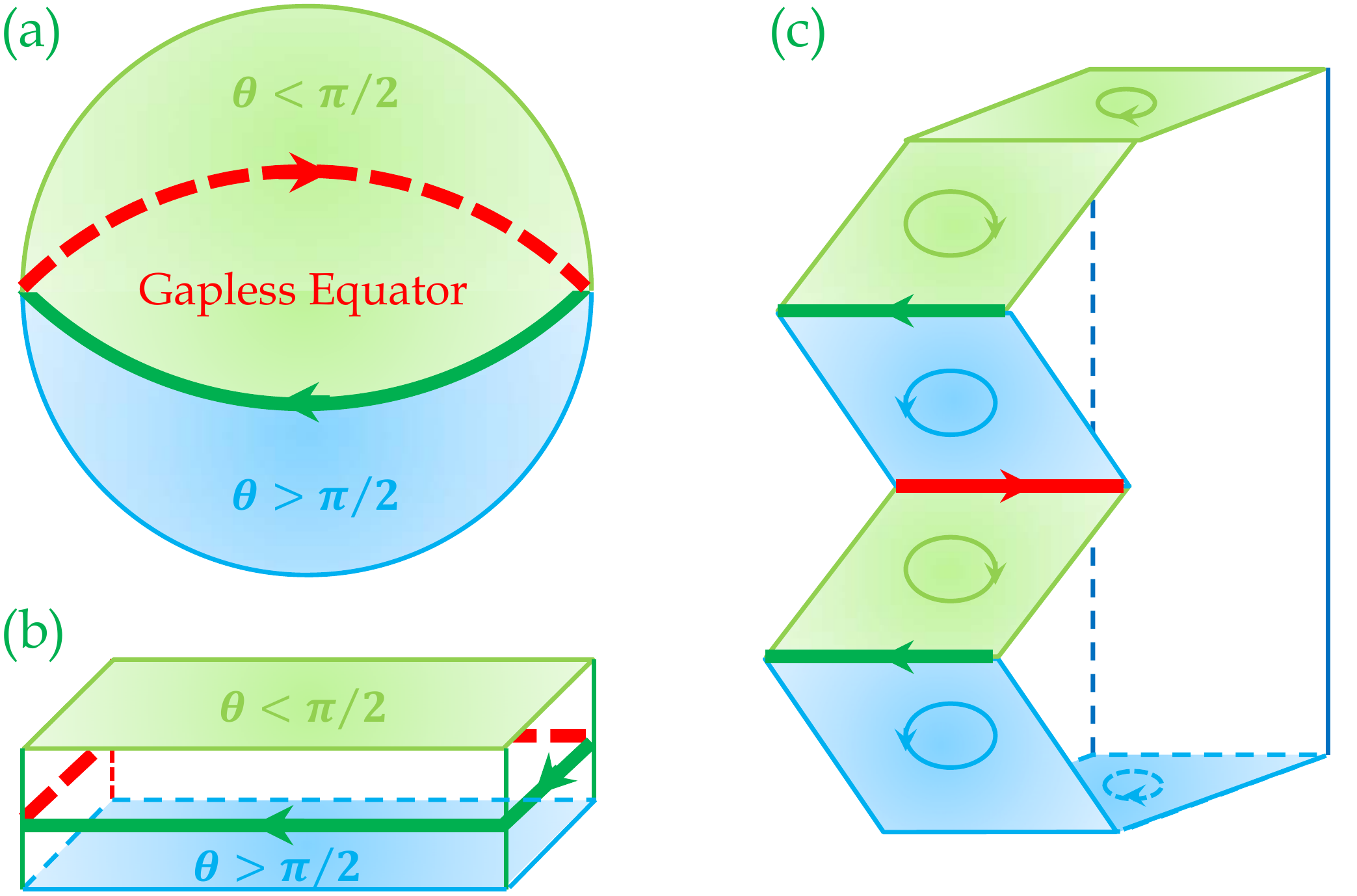}
}}
\caption{\label{fig:fg3} {(color online). Three TI shapes that support chiral edge states in the presence of a uniform exchange field
$\Delta_{\rm z}\sigma_{\rm z}$ on the surface. (a) A spherical TI, (b) a TI slab and (c) a TI with zigzag side faces. In all panels the
faces with $\theta<\pi/2$ ($\theta>\pi/2$) are colored in light green (cyan), and the chiral edge states are denoted by bold lines with arrows.}}
\end{figure}

{\color{cyan}{\indent{\em 1D chiral edge states.}}}---
We find that the momentum-space Berry curvature becomes nontrivial when $\mathcal{T}$ symmetry is broken by the mass
term $\alpha\Delta_{\rm z} S_2^{\rm z}$. For an arbitrary face, the Berry curvature reads
\begin{eqnarray}\label{eq:BC}
\Omega^{(\rm s)}_{\hat{k}_3}(\theta,{k}_1,{k}_{\rm y})=-\frac{s\alpha\Delta_{\rm z} v_1 v_{\shortparallel}}{2\varepsilon^3}\,,
\end{eqnarray}
where $\varepsilon=\sqrt{v_1^2 k_1^2+v_{\shortparallel}^2 k_{\rm y}^2+\alpha^2\Delta_{\rm z}^2}$ and the symbol $s=+(-)$
denotes the surface conduction (valence) band. The momenta are measured from the shifted DP given in Eq.(\ref{eqn:Hdp}) in the
presence of magnetization parallel to QL's. The {\em orbital} magnetic moment \cite{Niu,Zhang_SQH} carried by a surface state Bloch electron is
\begin{eqnarray}\label{eq:OM}
m^{(\rm s)}_{\hat{k}_3}(\theta,{k}_1,{k}_{\rm y})=-\frac{\alpha\Delta_{\rm z}m_{\rm e} v_1 v_{\shortparallel}}{\varepsilon^2}\mu_{\rm B}\,,
\end{eqnarray}
where $m_{\rm e}$ is the electron mass and $\mu_{\rm B}$ is the Bohr magneton. Unlike the Berry curvature, the orbital
magnetization is independent of the band index $s$. In the presence of an electric field in the surface plane, a
surface state electron acquires an anomalous transverse velocity proportional to the Berry curvature \cite{Niu,Zhang_SQH},
giving rise to an intrinsic Hall conductivity
\begin{eqnarray}
\label{eqn:Hall}
\sigma_{\rm H}=\frac{e^2}{2h}\left[\frac{\alpha\Delta_{\rm z}}{\varepsilon(E_{\rm F})}-{\rm sgn}(\alpha\Delta_{\rm z})\delta_{\rm s,+}\right]\,,
\end{eqnarray}
where $E_{\rm F}$ is the Fermi energy. Provided that $E_{\rm F}$ lies in the surface gap,
the surface band contributes $e^2/2h$ to the Hall conductivity, with the sign given by ${\rm sgn}(\alpha\Delta_{\rm z})$.

This Hall conductivity is half integer quantized but with opposite signs for crystal faces with surface normals that have
opposite $z$-projections ({\em i.e.} perpendicular to QL's), even though the surface magnetization is
uniform. Since $\Delta\sigma_{\rm H}=e^2/h$ across the interface, there must be a chiral edge state channel whenever there
is an edge or a narrow side face that connects
two faces whose surface normals have opposite $z$-projections.
This is the {\em criterion} for the existence of chiral edge states in the presence of uniform $\Delta_{\rm z}\sigma_{\rm z}$ magnetization on the surface of a TI.

We now propose three TI shapes that support chiral edge states in the presence of surface magnetization perpendicular to QL's and uniform on
all relevant faces. For a spherical TI, shown in Fig.\ref{fig:fg3}(a), the mass term and the Hall conductivity switch
sign across the equator ($\alpha=0$). Therefore, there is a chiral channel along the
equator for gapless edge states. A TI slab depicted in Fig.\ref{fig:fg3}(b) is topologically equivalent in shape to a spherical TI,
with the upper (lower) hemisphere becoming the top (bottom) cleavage surface. Similarly, there is a gapless chiral state along the
side faces when the exchange field effect dominates over the finite size effect. The QAH effect in this bilayer
(BL) system, studied before \cite{TFT,slab,Hughes} and a special case where our criterion applies, can be alternatively described in the crystal frame as follows:
\begin{eqnarray}
\label{eqn:bilayer}
\mathcal{H}_{\rm BL}=v_{\shortparallel}(k_{\rm y}\sigma_{\rm x}-k_{\rm x}\sigma_{\rm y})\tau_{\rm x}
-m_{\rm t}\tau_{\rm z}+\Delta_{\rm z}\sigma_{\rm z}\,,
\end{eqnarray}
where $\tau_{\rm x}=\pm$ respectively represent the bottom and top surfaces and $m_{\rm t}$ is a trivial mass due to
finite size tunneling between the surfaces. We further obtain the four-band energy dispersions:
\begin{eqnarray}
\varepsilon_{\rm BL}=\pm\sqrt{v_{\shortparallel}^2 k_{\shortparallel}^2+(m_{\rm t}\pm \Delta_{\rm z})^2}\,.
\end{eqnarray}
As the exchange field strength $\Delta_{\rm z}$ is turned up from zero, the energy gap closes at $\Delta_{\rm z}=\pm m_{\rm t}$
and reopens, indicating the topological distinction between the magnetization-induced gap with respect to the tunneling-induced gap.  Further analysis
using Eq.(\ref{eqn:Hall}) shows the two valence bands have a total $e^2/h$ contribution to $\sigma_{\rm H}$ when
$|\Delta_{\rm z}|>|m_{\rm t}|$, leading to a QAH effect. This analysis applies generally to any other slab with $\theta\neq \pi/2$.

Our criterion also predicts chiral edge states on a TI with a more remarkable shape, as depicted in Fig.\ref{fig:fg3}(c). In
the intergrowth with a ferromagnet \cite{Cava}, a TI often has zigzag side faces, typically a few microns in size. Each convex corner of
the zigzag side connects an upper face with $\theta_{\rm N}<\pi/2$ and a lower face with $\theta_{\rm S}>\pi/2$
while a concave corner connects the two faces upside down. As a consequence, the zigzag side exhibits staggered chiral corner
states with opposite velocities along the convex and concave corners. (These staggered chiral states can be coupled by interactions and may exhibit some exotic Luttinger liquid behaviors.) In the presence of an electric field perpendicular to the average face of the zigzag side, there will be a net chiral current carried by edge states at the convex or concave corners depending on the electric polarity, while for an electric field perpendicular to QL's edge states with opposite chiralities are both populated and the Hall currents become counterpropagating and canceling each other out on average.

In the limit of $\theta_{\rm N,S}\simeq\pi/2$, the surface state Hilbert spaces for the upper and lower faces are both the negative
eigenstate of $S_1^{\rm x}(\pi/2)=\sigma_{\rm y}\tau_{\rm y}$. The chiral corner states are also pseudospin filtered,
since they satisfy $S_2^{\rm x}(\pi/2)=\sigma_{\rm z}\tau_{\rm z}=\pm 1$ where the sign depends on the polarity of the
$\Delta_{\rm z}\sigma_{\rm z}$ magnetization. These two features are analogous to the case of a magnetic domain wall deposited on the cleavage surface, where the chiral edge state is not only orbital chiral but also spin filtered.

In the opposite limit in which $\theta_{\rm N}\simeq 0$ and $\theta_{\rm S}\simeq\pi$, these zigzag side faces become a chain
of staggered top and bottom cleavage surfaces, in which the two opposite edges of each face are joined respectively to the
opposite edges of the neighboring upper and lower faces. Our proposed criterion can be applied to each pair of neighboring
faces. The origin of their chiral corner states can be also intuitively understood by our previous analysis
(Eq.(\ref{eqn:bilayer})) designed for parallel double layers. In such a limit, the top and bottom surface states have zero
orbital overlap in the sense that they are negative and positive eigenstates of $\tau_{\rm x}$, respectively. Under uniform
$\Delta_{\rm z}\sigma_{\rm z}$ magnetization, the two surface valence bands contribute $e^2/h$ to $\sigma_{\rm H}$, however, the chiral edge
state is not spin filtered because of the opposite helicities at the two surfaces. These features are quite different from
the situation for two joined side faces or with a magnetic domain wall on the cleavage plane.
\begin{figure}[t]
\centering{
\scalebox{0.55} {\includegraphics*{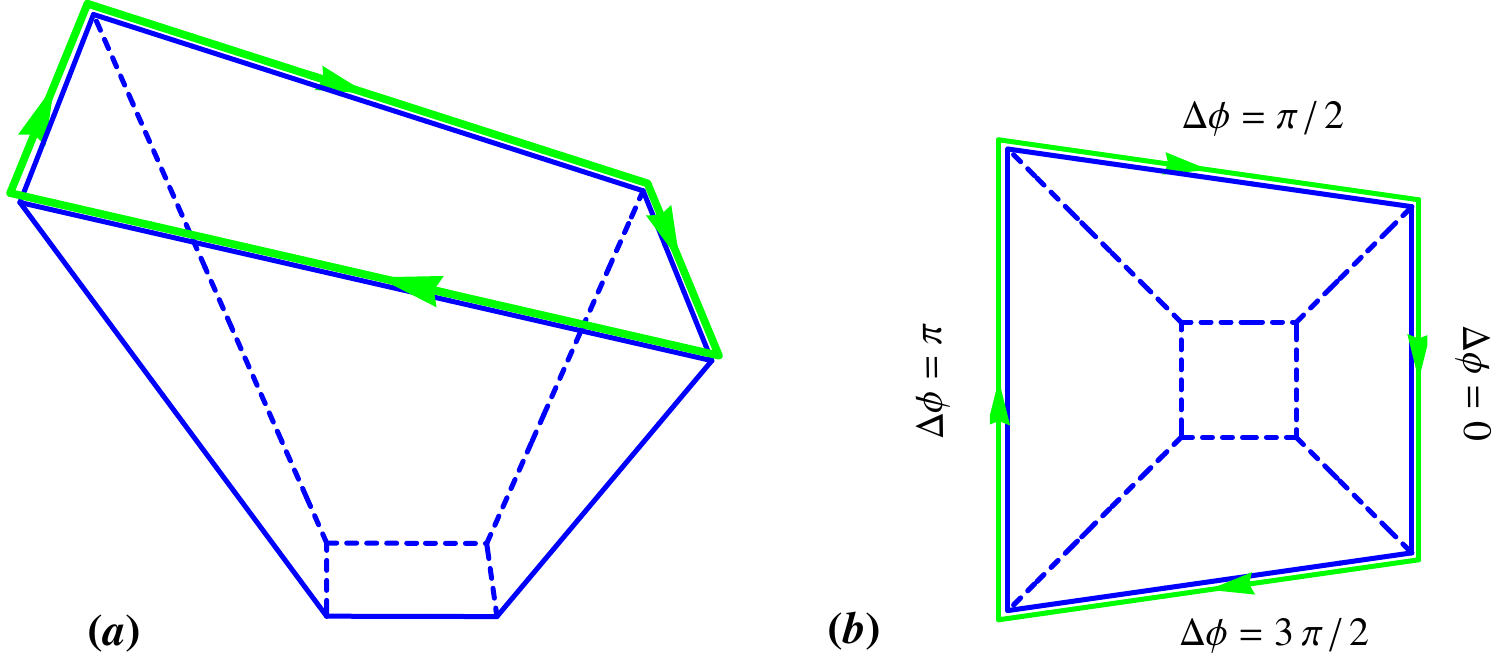}}}
\caption{\label{fig:fg4} {(color online).
(a) A schematic TI crystal that has a top face with $0<\theta_t<\pi/2$, a bottom face with $\theta_b=\pi$, and four surrounding faces with the same $\theta_s$ and $\pi/2<\theta_s<\pi$. (b) The top view of (a). When perpendicular magnetization is present, there is a chiral state along the edges of the top face but no one on the bottom edges.  The four surrounding faces have $\Delta\phi=\phi_s-\phi_t=0$, $\pi/2$, $\pi$, and $3\pi/2$, respectively.\cite{theta}
}}
\end{figure}

We now turn on the azimuthal angle, to study its influence on chiral edge states.
Since the bulk crystal has $\mathcal{C}_3$ symmetry along $\hat{k}_{\rm z}$ which upgrades to continuous rotational symmetry in linear order we can set $\phi=0$ along an arbitrary axis perpendicular to $\hat z$.
But more generally for two crystal faces joined at an edge with normals along {\em different} azimuthal angles, we need to specify
their difference $\Delta \phi$ to determine the chirality of their edge state. In fact,
the criterion for the existence of chiral edge state at a TI
corner can be relaxed to $\theta(\Sigma_{\rm \sigma})<\pi/2$ and $\theta(\Sigma_{\bar{\rm\sigma}})>\pi/2$ with
$\sigma=S\,\text{or}\,N$, where $\Sigma_{\rm S,N}(\theta,\phi)$ could even be curved faces or have different azimuthal angles.
Fig.\ref{fig:fg4} sketches a chiral state along the edges that
connect a top face with four surrounding faces that have the same $\theta_s$ but different $\phi_s$. The edge states have
opposite chiralities for $\Delta\phi=0$ and $\pi$ but not for other pair of azimuthal angles that differ in $\pi$.

{\color{cyan}{\indent{\em Discussions.}}}--- In conclusion, we provide a general theory that allows a thorough
understanding of the interaction of TI surface states with a ferromagnetically ordered medium, with a surprising criterion for the presence of a chiral
edge state (QAH effect) with no need of magnetic anisotropy.  This approach may also simplify the interferometry of
Majorana fermions \cite{MF_Kane,MF_Beenakker} that requires to generate chiral edge states on the TI surface.
Our proposed chiral edge states may be accessible by STM or in multi-terminal transport at the corners of a TI zigzag side that form \cite{Cava}
in the intergrowth with a ferromagnet. For a $\Delta\sim 1$ meV gap induced by a Zeeman exchange field, the coherence length $\ell_{\rm c}=\hbar v/\Delta$ is about a micron, which is comparable to the size of zigzag side face. Our present work may shed light on cleavage surface transport experiments where it is crucial to minimize the influence from the side face, {\em i.e.}, making samples in the square shape. On the other hand, this work also provides a new strategy for the fabrication of electronic devices that exploit the crystal face dependence of
TI surface state phenomena. We thus suggest reexamining anomalies in existing data by taking into account the dependence of the
surface states on the surface orientation and look forward to more explorations on the non cleavage surfaces.

A Zeeman field that couples to spin could also be introduced by doping
TI's with magnetic impurities \cite{dop1,dop2,slab,dopnew0,dopnew1,dopnew2,dopnew3,dop3,dop4,Samarth,dopNE,dopNomura,dopSamarth,dop7,dop8,dop9,dop10,dop11,dop_Yao,dop12,dop13}
or partially by applying a parallel magnetic field \cite{parallel-B,parallel_Zhang} instead of depositing
a ferromagnetic film \cite{Cava,Garate,Hughes,Wei} on the surface.
With these experimental progresses, QAH effects are likely to occur when their TI samples are fabricated in geometries similar to Fig.\ref{fig:fg3} or Fig.\ref{fig:fg4}.
We have noticed that a very special case of our proposed physics, {\em i.e.}, a QAH effect in a magnetic doped ${\rm Bi_2Se_3}$
slab with only two cleavage surfaces \cite{TFT,slab,Hughes} and thickness less than six QL's, has been demonstrated by the first principle calculations \cite{slab}.
While observing QAH effects in thin films is still experimentally challenging,
our newly discovered criterion does not necessarily require such a limited geometry and constitutes a significant advance.
Finally we point out that the surface magnetization can be built into the topological boundary condition of
TI's as a family of surface potentials \cite{TISS} that break $\mathcal{T}$ symmetry but preserve $\mathcal{P}$ symmetry.

This work has been supported by DARPA under grant SPAWAR N66001-11-1-4110.


\begin{thebibliography}{100}
\bibitem{notation}
 The effective Hilbert space near the bulk gap is spanned by states with angular momentum (proportional to the electron spin) $\sigma=\pm 1/2$ and $p_{\rm z}$ orbitals $\tau_{\rm z}=\pm 1$. The $\tau_{\rm z}=1\;(-1)$ orbital is dominated by states from Bi (Se) atoms, due to the large energy difference between $4p$ (Se) and $6p$ (Bi) principal quantum levels. These two states are inverted at $\Gamma$ point due to spin-orbital couplings.
\bibitem{g-factor}
 This estimation has to be determined by experiments on the surface states of relevant materials. If the $\sigma_{\rm x}\tau_{\rm z}$ field turns out to be comparable to the $\sigma_{\rm x}$ field, our results throughout the paper still holds by turning off $\Delta_{\rm x}$.
\bibitem{overlap}
 We note that in this case $\Delta_{\rm x,z}$ are reduced, taking into account the overlap of surface states on opposite faces in the surface normal direction.
\bibitem{theta}
 We have chosen to illustrate the case $\theta_{\rm t}+\theta_{\rm s}<\pi$ in this figure. The case of $\theta_{\rm t}+\theta_{\rm s}>\pi$ or a concave side corner can be considered in a similarly way. For the special case when $\theta_{\rm t}+\theta_{\rm s}=\pi$ and $\Delta\phi=\pi$, a bilayer model similar to Eq.(\ref{eqn:bilayer}) applies here.
\end{thebibliography}
\end{document}